\documentclass[preprint,5p,times,sort&compress,twocolumn]{elsarticle}

\usepackage{amsthm, amsfonts, amssymb, comment}	
\usepackage[fleqn]{amsmath}				
\usepackage{graphicx}					
\usepackage{txfonts}					
\usepackage{hyperref,xcolor}			
\usepackage{geometry}                   
\usepackage{lineno}						
\modulolinenumbers[1]					
\usepackage{mathtools, cuted}           

\newcommand{\newc}{\newcommand}
\newc{\beq}{\begin{equation}}
\newc{\eeq}{\end{equation}}
\newc{\bea}{\begin{array}}
	\newc{\eea}{\end{array}}
\newcommand{\ben}{\begin{eqnarray}}
\newcommand{\een}{\end{eqnarray}}
\newc{\ra}{\rightarrow}
\newc{\bfx}{{\bf x}}
\newc{\bfV}{{\bf V}}
\newc{\cO}{{\cal O}}
\newc{\bfv}{{\bf v}}
\newc{\bfu}{{\bf u}}
\newc{\bfp}{{\bf p}}
\newc{\ve}{{\varepsilon}}
\newc{\Psibar}{\overline\Psi}
\newc{\w}{{\bf w}}
\newc{\E}{{\mathbf{E}}}
\newc{\EE}{{\mathcal E}}
\newc{\bfn}{{\mathbf\nabla}}
\newc{\la}{{\cal L}}
\newc{\tla}{{\tilde{\cal L}}}
\newc{\bp}{{\bf p}}
\newc{\ho}{\hookrightarrow }
\newc{\bP}{{\bf P}}
\newc{\pd}{{\partial}}
\newc{\piv}{{\partial_4}}
\newc{\pv}{{\partial_5}}
\newc{\bJ}{{\bf J}}
\newc{\bze}{{\mathbf 0}}
\newc{\bK}{{\bf K}}
\newc{\tphi}{{\tilde\phi}}
\newc{\tF}{{\tilde F}}
\newc{\tD}{{\tilde D}}
\newc{\tJ}{{\tilde J}}
\newc{\tj}{{\tilde j}}
\newc{\bD}{{\bf D}}
\newc{\tvphi}{{\tilde\varphi}}
\newc{\trho}{{\tilde\rho}}
\newc{\ttheta}{{\tilde\theta}}
\newc{\tpsi}{{\tilde\psi}}
\newc{\tu}{{\tilde u}}
\newc{\cD}{{\cal D}}
\newc{\tPhi}{{\tilde\Phi}}
\newc{\tPsi}{{\tilde\Psi}}
\newc{\tA}{{\tilde A}}
\newc{\talpha}{{\tilde\alpha}}
\newc{\tbeta}{{\tilde\beta}}
\newc{\bA}{{\mathbf A}}
\newc{\bB}{{\bf B}}
\newc{\br}{{\bf r}}
\newc{\sig}{{\mathbf\sigma}}
\newc{\eg}{{\rm e.g.\ }}
\newc{\ie}{{\rm i.e.\ }}
\newcommand{\bey}{\begin{eqnarray}}
\newcommand{\pslash}{\not{\hbox{\kern-2.3pt $p$}}}
\newcommand{\pdslash}{\not{\hbox{\kern-2pt $\partial$}}}
\newcommand{\eey}{\end{eqnarray}}

\begin{document}
	
	\begin{frontmatter}
		\title{Splitting frequency of the $(2+1)$-dimensional Duffin-Kemmer-Petiau oscillator in
			an external magnetic field}
		
		\author[salvador]{Ignacio S. Gomez\corref{cor1}}
		\ead{nachosky@fisica.unlp.edu.ar}
		\author[salvador]{Esdras S. Santos}
		\ead{esdras.santos@ufba.br}
		\author[salvador]{Olavo Abla}
		\ead{olavo.abla@ufba.br}
		\cortext[salvador]{Corresponding author}
		\address[salvador]{Instituto de F\'{i}sica, Universidade Federal da Bahia,
			Rua Barao de Jeremoabo, 40170-115 Salvador-BA, Brazil}
		

		\begin{abstract}
			We revisit the (2+1)-dimensional DKP oscillator in an external magnetic field by means of 4$\times$4 and 6$\times$6 representations of the DKP field, thus obtaining several cases studied in the literature. We found an splitting in the frequency of the DKP oscillator according to the spin projection that arises as an interplay between the oscillator, the external field and the spin, from which the energies and the eigenfunctions are expressed in a unified way. For certain critical values of the magnetic field the oscillation in the components of spin projections -1 and 1 is cancelled. We study the thermodynamics of the canonical ensemble of the vectorial sector, where a phase transition is reported when the cancellation of the oscillation occurs. Thermodynamic potentials converge rapidly to their asymptotical expressions in the high temperature limit with the partition function symmetric under the reversion of the magnetic field.
		\end{abstract}

		\begin{keyword}
			DKP oscillator \sep splitting frequency
			\sep relativistic wave equations \sep vectorial sector thermodynamics
			
		\end{keyword}
		
	\end{frontmatter}
	
	\nolinenumbers   
	
	\section*{Introduction}
	
	The Duffin-Kemmer-Petiau (DKP) equation is a relativistic first-order wave equation which
	describes scalar and vector fields through a unified formalism \cite{Duffin,Kemmer,Petiau,Nieto}.
	Using its representations \cite{Corson,representations} or projectors \cite{Umezawa,remarks} as well as a rich variety of couplings \cite{Guertin} for the spin zero and spin one sectors, this theory has been applied to the study of many different problems: the
	scattering of $K^{+}$ nucleus \cite{knucleus}, quantum chromodynamics \cite{Gribov}, studies on the S-matrix \cite{Pimentel-1}, causality of the DKP theory \cite{Pimentel-2}, minimal and non-minimal coupling in a general representation of DKP matrices \cite{Luciano}, Bose-Einstein condensation \cite{Pimentel-3,Galileano-4}, breaking of Lorentz symmetry \cite{LIV-DKP}, curved space–time \cite{Pimentel-4}, Aharonov-Bohm potential \cite{Boumali-4}, studies of the phase in Aharonov-Casher effect \cite{Swansson}, Galilean five-dimensional formalism \cite{Galileano-4,Galileano-1,Galileano-2,Galileano-3} and several others applications \cite{Nedjadi1,Boumali-5,Vianna2,Boutabia,Ghose,Pimentel-5,Hassanabadi1,Mirza,DKP-Osc,Hassanabadi2,Boumali-1,DKP-KGPROCA,Wu,DKPNC,DKPMAG}.
	
	Particularly, in \cite{Dirac-Osc} the authors have shown that the (2+1)-dimensional Dirac oscillator with frequency $\omega$ in an external magnetic field taken along the $z$-direction and represented by the vector potential ${\bf A}=\frac{B}{2}\left(-y,\;x\right)$, can be mapped onto the Dirac oscillator without magnetic field but with reduced angular frequency given by  $\omega-\widetilde{\omega}$, where $\widetilde{\omega}=\frac{eB}{2mc}$ and $e$ and $m$ are the charge and the mass of the electron. This result has been used, for example, to study atomic transitions  in a radiation field \cite{Dirac-Osc} as well as for
its corrections through the generalized uncertainty principle \cite{DO+GUP}. Following the same approach,
	the (2+1)-dimensional DKP oscillator (DKPO) has been studied under an external magnetic field \cite{Mirza,Boumali-1,Hassanabadi2,DKPMAG}, where the authors have calculated the eigensolutions of massive spin-0 and spin-1 particles both in commutative and non-commutative phase-space.

	In this letter we revisit this problem in commutative space in order to show some interesting properties in the spin
    one sector, which can be found with suitable conditions for components of the DKP field.
	The work is organized as follows.
	We begin by considering a
	scalar $4\times 4$ and a $6\times 6$ representations
	for the
	DKP field in the presence of an external magnetic field, that allows to study several cases of interest in the literature. Then, we analyze the scalar and vectorial cases as well as some special ones, thus obtaining the simplified DKPO and the two-dimensional DKPO along with their
    energies, eigenfunctions
	and degeneracies. In the spin one sector of the DKP representation we found a relationship (splitting)
	between the frequency of the DKP oscillator
	and the projection spin
	of the particles, whose main effects are to flip the vectorial components $s_{1,2}=\pm1$
	when the direction of the magnetic field is reversed and to cancel the oscillation
	in the $s_{1,2}$ component for some values of the magnetic field.
	Next, we study the statistical properties of the canonical ensemble of the vectorial sector
	and we derive its thermodynamics, exhibiting
	phase transitions for some values of the magnetic field, as a consequence of the splitting.
	Finally, we summarize the results and draw the conclusions.
	\section*{DKP equation and some representations}
	
	The DKP equation is written by the form \cite{Duffin}-\cite{Petiau} (see the beautiful review by Krajcik and Nieto \cite{Nieto} and the references therein)
	\bey
	\left(i\hbar \beta^\mu \partial_\mu-mc\right)\Psi=0,\label{DKP}
	\eey
	where $\Psi$ is a DKP field of mass $m$ and $\beta$-matrices satisfy the DKP algebra
	\begin{equation}\label{DKPalgebra}
	\beta^{\mu}\beta^{\lambda}\beta^{\nu}+\beta^{\nu}\beta^{\lambda}\beta^{\mu}=\beta^{\mu}\eta^{\lambda\nu}+\beta^{\nu}\eta^{\lambda\mu}.
	\end{equation}
	
	As predicted by \cite{Corson}  and showed in \cite{representations}, the DKP equation in $(2+1)$-dimensional space-time has two representations: a $4\times 4$ for the scalar sector and a $6\times 6$ for the vector sector, with the Lorentz metric $g^{\mu\nu}$ given by $g^{00}=-g^{11}=-g^{22}=1$. The DKP representation for the scalar sector is given by matrices
	\bey
	&&\beta^0=
	\left(\begin{array}{cccc}
		\cdot&\cdot&\cdot&1\\
		\cdot&\cdot&\cdot&\cdot\\
		\cdot&\cdot&\cdot&\cdot\\
		1&\cdot&\cdot&\cdot
	\end{array}\right),
	\beta^1=
	\left(\begin{array}{cccc}
		\cdot&\cdot&\cdot&\cdot\\
		\cdot&\cdot&\cdot&1\\
		\cdot&\cdot&\cdot&\cdot\\
		\cdot&-1&\cdot&\cdot
	\end{array}\right),\nonumber\\
	&&\beta^2=
	\left(\begin{array}{cccc}
		\cdot&\cdot&\cdot&\cdot\\
		\cdot&\cdot&\cdot&\cdot\\
		\cdot&\cdot&\cdot&1\\
		\cdot&\cdot&-1&\cdot
	\end{array}\right),
	\label{s-1}
	\eey
	and the DKP representation for the vector sector is given by the $6\times 6$ matrices
	\bey
	\beta^0=
	\left(\begin{array}{cccccc}
		\cdot&\cdot&\cdot&-1&\cdot&\cdot\\
		\cdot&\cdot&\cdot&\cdot&-1&\cdot\\
		\cdot&\cdot&\cdot&\cdot&\cdot&\cdot\\
		-1&\cdot&\cdot&\cdot&\cdot&\cdot\\
		\cdot&-1&\cdot&\cdot&\cdot&\cdot\\
		\cdot&\cdot&\cdot&\cdot&\cdot&\cdot
	\end{array}\right),\;\;\nonumber
	\eey
	\bey
	\beta^1=
	\left(\begin{array}{cccccc}
		\cdot&\cdot&\cdot&\cdot&\cdot&1\\
		\cdot&\cdot&\cdot&\cdot&\cdot&\cdot\\
		\cdot&\cdot&\cdot&\cdot&1&\cdot\\
		\cdot&\cdot&\cdot&\cdot&\cdot&\cdot\\
		\cdot&\cdot&-1&\cdot&\cdot&\cdot\\
		-1&\cdot&\cdot&\cdot&\cdot&\cdot
	\end{array}\right),\nonumber
	\eey
	\bey
	\beta^2=
	\left(\begin{array}{cccccc}
		\cdot&\cdot&\cdot&\cdot&\cdot&\cdot\\
		\cdot&\cdot&\cdot&\cdot&\cdot&1\\
		\cdot&\cdot&\cdot&-1&\cdot&\cdot\\
		\cdot&\cdot&1&\cdot&\cdot&\cdot\\
		\cdot&\cdot&\cdot&\cdot&\cdot&\cdot\\
		\cdot&-1&\cdot&\cdot&\cdot&\cdot
	\end{array}\right).\label{v-1}
	\eey

	The DKPO in the presence of an external magnetic field ${\bf A}=\frac{B}{2}\left(-y,\;x\right)$ is performed by the coupling \cite{Mirza,Boumali-1,Hassanabadi2,DKPMAG}
	\bey
	{\bf p}\rightarrow {\bf p}-\frac{q}{c}{\bf A}-im\omega\eta^0{\bf r},\label{s-2}
	\eey
	where $\omega$ is the angular frequency of the oscillation, $q$ is the charge of the boson  and $\eta^0=2\left(\beta^0\right)^2-\mathbf{1}$. The coupled DKP equation is written by
	\bey
\left[-\beta^0 E+c\beta\cdot\left({\bf p}-\frac{q}{c}{\bf A}-im\omega\eta^0{\bf r}\right)+mc^2\right]\Psi=0,\label{s-3}
\eey
where we have used $-i\hbar\partial_\mu=-p_\mu=\left(\frac{E}{c},\;{\bf p}\right)$.

\section*{Scalar DKPO in a magnetic field}
 For the scalar sector, using a four components $\Psi^T=\left(\Psi_0,\;\Psi_1,\;\Psi_2,\;\Psi_3,\;\right)$ and $\widetilde{\omega}=\frac{qB}{2mc}$, the coupled equation (\ref{s-3}) provides the following four equations
\bey
&&mc^2\Psi_0=E\Psi_3\label{DKP-s-1}\\
&&mc^2\Psi_1=-c\left(p_x+m\widetilde{\omega}y-im\omega x\right)\Psi_3\label{DKP-s-2}\\
&&mc^2\Psi_2=-c\left(p_y-m\widetilde{\omega}x-im\omega y\right)\Psi_3\label{DKP-s-3}\\
&&mc^2\Psi_3=E\Psi_0+c\left(p_x+m\widetilde{\omega}y+im\omega x\right)\Psi_1\nonumber\\
&&+c\left(p_y-m\widetilde{\omega}x+im\omega y\right)\Psi_2.\label{DKP-s-4}
\eey
Performing the combination of these equations we have
\bey
\left(E^2-m^2c^4\right)\Psi_3=c^2({\bf p}^2+m^2 \omega_0^2 {\bf r}^2
-2m\hbar \omega-2m\widetilde{\omega} L_z)\Psi_3,\label{s-4}
\eey
where $\omega_0=\sqrt{\omega^2+\widetilde{\omega}^2}$, ${\bf r}^2=x^2+y^2$ and $L_z=xp_y-yp_x$ is the z-component of the angular momentum.  This result is in agreement with \cite{Hassanabadi2},
for a null non-commutative factor. Thus, following the standard approach, the energy spectrum is given by
\bey
E_{n,l}=\pm mc^2\sqrt{1-\frac{2\hbar(|l|\widetilde{\omega}+\omega)}{mc^2}
		+\frac{\hbar\omega_0}{mc^2}(4n+2(|l|+1))}, \label{scalar-DKPO-spectrum}
\eey
where $n=0,1,2,3,...$ is the quantum number associated to the energy and $l=0,\pm 1,\pm 2,\pm 3...$ is the angular momentum quantum number.
The eigenfunction $\Psi_3$ is calculated using the polar coordinates ($r,\;\theta$), we have
\bey
\left(\Psi_3\right)_{n,l}=\left(\frac{m\omega_0}{\hbar} r^2\right)^{|l|/2} e^{il\theta}e^{-\frac{m\omega_0}{2\hbar} r^2} L^{|l|}_n \left(\frac{m\omega_0}{\hbar} r^2\right),\label{scalar-DKPO-solution}
\eey
where $L_n$ stands for the $nth$ Laguerre polynomial and the non-relativistic limit of \eqref{scalar-DKPO-spectrum} is calculated taking $E=\epsilon+mc^2$ with $\epsilon<<mc^2$ giving $\left(E^2-m^2c^4\right)/2mc^2\cong\varepsilon$. The others components of the
DKP eigenfunction, $\Psi$, can be easily calculated using the expressions (\ref{DKP-s-1})-(\ref{DKP-s-3}).
These
results show a perfect equivalence between the representations formalism \cite{representations} and the one employed on the magnetic DKP oscillator from the past years \cite{Mirza,Boumali-1,Hassanabadi2,DKPMAG}, considering the scalar sector and the commutative phase-space. Next, we want to show the advantage of this formalism, with no particular choice to describe the system like was made before, where
the expected results are the spin effects and a curious ``splitting" on the frequencies.


\section*{Vector DKPO in a magnetic field: splitting in the frequency}

For the vectorial sector, using the equation ($B3$) of \cite{Luciano} and taking $R^0\Psi=0$ (thus making the scalar component equal to zero), the equation ($49$) of \cite{Luciano} is simplified by cancelling the term proportional to $1/m^2$. In this way it is obtained a simplified equation for the ($3+1$)-dimensional vector DKP oscillator (SDKPO), which is composed by the usual three-dimensional oscillator added to a spin-orbit coupling. Following this reasoning, in the $(2+1)$-dimensional case we choose conveniently the six components of the DKP field $\Psi^T=\left({\bf a},\;b,\;{\bf d},\;0\right)$ where ${\bf a}=\left(a_1,\;a_2\right)$ and ${\bf d}=\left(d_1,\;d_2\right)$. The null component was chosen in order to obtain the ($2+1$)-dimensional SDKPO. Using $\Psi$, from the representation (\ref{v-1}) as well as the coupling (\ref{s-3}) in the (\ref{DKP}), we obtain the six equations
\bey
mc^2 a_1=-Ed_1\label{1}\\
mc^2 a_2=-Ed_2\label{2}\\
mc^2b=-c\left(p_x+m\widetilde{\omega}y-im\omega x\right)d_2\nonumber\\
+c\left(p_y-m\widetilde{\omega}x-im\omega y\right)d_1\\
-mc^2 d_1=Ea_1+c\left(p_y-m\widetilde{\omega}x+im\omega y\right)b \\
-mc^2 d_2=Ea_1-c\left(p_x+im\widetilde{\omega}y+im\omega x\right)b\\
0=c\left(p_x+m\widetilde{\omega}y-im\omega x\right)a_1-\nonumber\\
c\left(p_y-m\widetilde{\omega}x-im\omega y\right)a_2.
\eey
By substituting Eqns. (\ref{1}) and (\ref{2})
into the others, we have the equations for the components $b,\;d_1$ and $d_2$, given by
\bey
&&(E^2-m^2c^4)b=c^2(p_x+m\widetilde{\omega}y-im{\omega}x)(p_x+im\widetilde{\omega}y+im\omega x)\nonumber\\
&&+c^2(p_y-m\widetilde{\omega}x-im\omega y)(p_y-m\widetilde{\omega}x+im\omega y)]b\label{escalar}\\
&&(E^2-m^2 c^4) d_1=-c^2(p_y-m\widetilde{\omega}x+im\omega y)\times\nonumber\\
&&[(p_x+m\widetilde{\omega}y-im\omega x)d_2-(p_y-m\widetilde{\omega}x-im\omega y)d_1]  \label{v-2}\\
&&(E^2-m^2 c^4)  d_2= c^2(p_x+m\widetilde{\omega}y+im\omega x)\times\nonumber\\
&&[(p_x+m\widetilde{\omega}y-im\omega x)d_2-(p_y-m\widetilde{\omega}x-im\omega y)d_1] \label{v-3}\\
&&0=(p_x+m\widetilde{\omega}y-im\omega x)d_1+(p_y-m\widetilde{\omega}x-im\omega y)d_2.
\label{v-4}
\eey
Replacing \eqref{v-2} and (\ref{v-3}) in (\ref{escalar}) it follows that
\ben
\left(E^2-m^2 c^4\right) b=
c^2\left[{\bf p}^2+m^2 \omega_0^2{\bf r}^2+2m\hbar \omega-2m\widetilde{\omega}L_z\right]b,\label{escalar1}
\een
where $\omega_0=\sqrt{\omega^2+{\widetilde{\omega}}^2}$. The expression of above show us that the component $b$, like (\ref{s-4}),
behaves as a scalar DKPO in a magnetic field. The energy spectrum for (\ref{escalar1}) is obtained by the exchange $\omega\rightarrow -\omega$ in (\ref{scalar-DKPO-spectrum}), while its eigenfunction is equal to (\ref{scalar-DKPO-solution}). In other hand, if we apply the operator $c^2\left(p_x+m\widetilde{\omega}y+im\omega x\right)$ to (\ref{v-4}) and add the result to (\ref{v-2}), and then we apply the operator $c^2\left(p_y-m\widetilde{\omega}x+im\omega y\right)$ to (\ref{v-4}) and add the result to (\ref{v-3}), we have
\bey
\left(E^2-m^2 c^4\right)d_1=c^2\bigg[{\bf p}^2+m^2\left(\omega^2+{\widetilde{\omega}}^2\right){\bf r}^2-2m\hbar\omega\nonumber\\
-2m\widetilde{\omega}L_z\bigg]d_1+ic^2\left(2m\hbar\widetilde{\omega}+2m\omega L_z-2m^2\omega\widetilde{\omega}{\bf r}^2\right)d_2\label{v-5}
\eey
and
\bey
\left(E^2-m^2 c^4\right) d_2=c^2\bigg[{\bf p}^2+m^2\left(\omega^2+{\widetilde{\omega}}^2\right){\bf r}^2-2m\hbar\omega\nonumber\\
-2m\widetilde{\omega}L_z\bigg]d_2-ic^2\left(2m\hbar\widetilde{\omega}+2m\omega L_z-2m^2\omega\widetilde{\omega}{\bf r}^2\right)d_1 \label{v-6}.
\eey
Thus, defining $d_1$ and $d_2$ by
\bey
d_1&=& \frac{\varphi_{2}+\varphi_{1}}{2}\label{v-7}\\
d_2&=& \frac{\varphi_{2}-\varphi_{1}}{2i},\label{v-8}
\eey
it follows that the equations (\ref{v-5})-(\ref{v-6}) can be rewritten by the form
\bey
\left(E^2-m^2c^4\right)\varphi_{1}=c^2\bigg[{\bf p}^2+m^2\left(\omega+\widetilde{\omega}\right)^2{\bf r}^2
\nonumber\\
-2m\hbar\left(\omega+\widetilde{\omega}\right)-2m\left(\omega+\widetilde{\omega}\right)L_z\bigg]\varphi_{1}
\label{v-9}
\eey
and
\bey
(E^2-m^2 c^4)\varphi_{2}=c^2\bigg[{\bf p}^2+m^2(\omega-\widetilde{\omega})^2{\bf r}^2
\nonumber\\
-2m\hbar (\omega-\widetilde{\omega})+2m(\omega-\widetilde{\omega})L_z\bigg]\varphi_{2}.
\label{v-10}
\eey
Equations (\ref{v-9}) and (\ref{v-10}) can be rewritten in the
more compact form
\bey
\left(E^2-m^2 c^4\right)\varphi_{i}=c^2\left({\bf p}^2+m^2\omega_{i}^2{\bf r}^2-2m\hbar
\omega_{i}-2m\omega_{i}L_z S_z\right)\varphi_{i},
\label{v-11}
\eey
where $i=1,2$ with $\omega_{1}=\omega+\widetilde{\omega}$, $\omega_{2}= \omega-\widetilde{\omega}$ and
$S_z\varphi_i=\hbar s_i\varphi$ with $s_{1}=-s_{2}=1$.
The last term $2\omega L_z$ in the right side of
(\ref{v-11}) plays the role of a spin-orbit term.
Following this reasoning, we can interpret the functions $\varphi_{i}$ as components of a vector $(\varphi_{1},\;\varphi_{2})$ whose spin projections along the $z$-direction are $+1$ and $-1$, respectively. Thus, in the presence of an external magnetic field, the spin one DKPO presents a scalar component, $b$, and two vector components, $\varphi_{1}$ and $\varphi_{2}$, which oscillate with the angular frequencies $\omega_0$, $\omega_{1}$  and $\omega_{2}$, respectively.
\begin{figure}
\includegraphics[width=\linewidth]{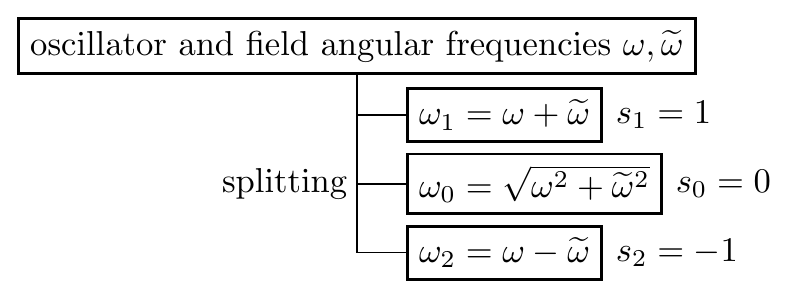}
\caption{\label{fig:diagram}
  Schematic representation of the frequency splitting of the DKPO
  (Eq. (\ref{v-11}))
  in the presence of an external magnetic field, according to
  the spin $s_i$ with $i=0,1,2$.
  }
\end{figure}
The expression (\ref{v-11}) has the form of the equation for the $($2+1$)$-dimensional SDKPO whose energy spectrum and eigensolutions are
\bey
E_{n,l}^{(i)}=\pm  mc^2\sqrt{1+\frac{\hbar \omega_i}{mc^2}\left(4n+2|l|\left(1-s_i\right)+2s_i\right)} \quad i=1,2
\label{spectrum-v}
\eey
where $n=0,1,2,3,...$, $l=0,\pm 1,\pm 2,\pm 3...$ with
$E_{n,l}^{(1)}$ and $E_{n,l}^{(2)}$ the energies associated to the angular 
frequencies $\omega_{1}$ and $\omega_{2}$, respectively. The eigenfunctions $\varphi_{i}$ are expressed by
\bey
\varphi_{n,l}^{(i)}=\left(\frac{m\omega_{i}}{\hbar} r^2\right)^{|l|/2} e^{il\theta}e^{-\frac{m\omega_{i}}{2\hbar} r^2} L^{|l|}_n \left(\frac{m\omega_{i}}{\hbar} r^2\right),\label{solution-v}
\eey
whose non-relativistic limit is calculated by the same way as pointed out previously.
The others components of the spin one DKP eigenfunction, $\Psi$, can be easily calculated using the expressions (\ref{DKP-s-1})-(\ref{DKP-s-3}),
whose non-relativistic limit is obtained taking $E=\epsilon+m^2c^4$ with $\epsilon<<mc^2$.

\section*{Special cases}
Below we explore the special cases associated to the situations where $d_1=0$, $d_2=0$, $d_1=\pm d_2$, $\omega=\pm \widetilde{\omega}$.
The angular frequency of the oscillator $\omega$ is assumed to be positive while the magnetic field
frequency $\widetilde{\omega}=\frac{qB}{2mc}$ can be positive or negative depending on the sign of $qB$.
For $\widetilde{\omega}>0$ (resp. $<0$) it follows that $\omega_2 < \omega_0 < \omega_1$ (resp. $\omega_1 < \omega_0 < \omega_2$) which shows that
the change of sign of $\widetilde{\omega}$ flips the frequencies of the vectorial sector.

\textbf{Cases  $d_1=0$ or $d_2=0$}

In these cases we consider the equations (\ref{v-5})-(\ref{v-6}) in order to
obtain: $\varphi_{1}=-\varphi_{2}$ for $d_1=0$ or $\varphi_{1}=\varphi_{2}$ for $d_2=0$. Hence, for both cases we have
\bey
\left(E^2-m^2 c^4\right) \varphi_{1,2}=
c^2\left[{\bf p}^2+m^2\omega_{0}^2{\bf r}^2-2\hbar \omega-2\widetilde{\omega}L_z \right]\varphi_{1,2}.
\label{v-12}
\eey
These cases show us the each component $\varphi_{1,2}$ of the eigenfunction behaves as scalar DKPO in the presence of the magnetic field, with angular frequency $\omega_0=\sqrt{\omega^2+{\widetilde{\omega}}^2}$. Consequently, this situation presents the same results
as pointed out in (\ref{scalar-DKPO-spectrum})-(\ref{scalar-DKPO-solution}) 
including the non-occurrence of the Zitterbewegung frequency. This result is in complete agreement with \cite{Boumali-1}.

\textbf{Case $d_1=id_2$}

In this case we consider (\ref{v-5})-(\ref{v-6}) in order to obtain $\varphi_{1}=0$ and
\bey
\left(E^2-m^2 c^4\right)\varphi_{2}=
c^2\left[{\bf p}^2+m^2\omega_{2}^2{\bf r}^2-2\hbar \omega_{2}+2\omega_{2}L_z\right]\varphi_{2}.
\label{v-13}
\eey
Thus, the problem is mapped onto DKPO without magnetic field but with
a reduced frequency $\omega_{2}=\omega-\widetilde{\omega}$ and
a spin projection $S_{z}=-1$.

\textbf{Case $d_1=-id_2$}

In this case we consider (\ref{v-5})-(\ref{v-6}) in order to obtain $\varphi_{2}=0$ and
\bey
\left(E^2-m^2 c^4\right) \varphi_{1}=
\left[{\bf p}^2+m^2\omega_{1}^2{\bf r}^2-2\hbar \omega_{1}-2\omega_{1}L_z\right]\varphi_{1}.
\label{v-14}
\eey
Thus, the problem is mapped onto DKP oscillator without magnetic field but
with an increased frequency $\omega_{1}=\omega+\widetilde{\omega}$ and
a spin projection $S_{z}=+1$.

\textbf{Case $\widetilde{\omega}=\omega$}

In this case we just need make $\omega=\widetilde{\omega}$ in the (\ref{v-11}),
thus obtaining
\bey
&\left(E^2-m^2c^4\right)\varphi_1=c^2\left({\bf p}^2+4m^2\omega^2{\bf r}^2-4\hbar\omega-4\omega L_z S_z\right)\varphi_1,\label{v-15}\\
&\left(E^2-m^2 c^4\right)\varphi_2=c^2{\bf p}^2\varphi_2.\label{v-15}
\eey
Hence, the problem is mapped onto DKP oscillator without magnetic field but with
a duplicated
frequency $2\omega$ and a spin projection $S_{z} =+1$ for the component $\varphi_{1}$, while
the component $\varphi_{2}$ stops the oscillation and behaves as a free  particle.

\textbf{Case $\widetilde{\omega}=-\omega$ }

In this case we just need make $\omega=-\widetilde{\omega}$ in the (\ref{v-11}), obtaining
\bey
&\left(E^2-m^2c^4\right)\varphi_2=c^2\left({\bf p}^2+4m^2\omega^2{\bf r}^2-4\hbar\omega-4\omega L_z S_z\right)\varphi_2,\label{v-15}\\
&\left(E^2-m^2 c^4\right)\varphi_1=c^2{\bf p}^2\varphi_1,\label{v-16}
\eey
Hence, the problem is mapped onto DKP oscillator without magnetic field but with
a duplicated
frequency $2\omega$ and a spin projection $S_{z} =-1$ for the component $\varphi_{2}$, while the
component $\varphi_{1}$ stops the oscillation and behaves as a free particle.
\begin{figure}[bt]
\centering
\begin{minipage}[h]{0.8\linewidth}
\includegraphics[width=1\linewidth]{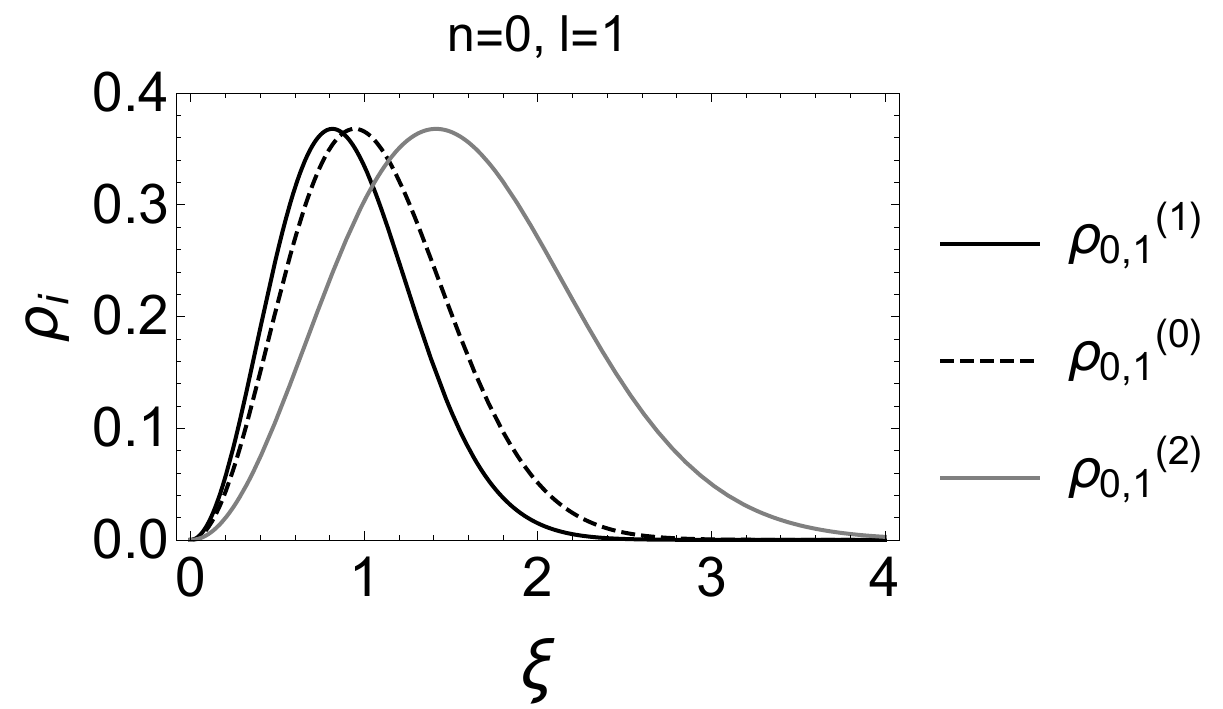}
\end{minipage}\\
\begin{minipage}[h]{0.8\linewidth}
\includegraphics[width=1\linewidth]{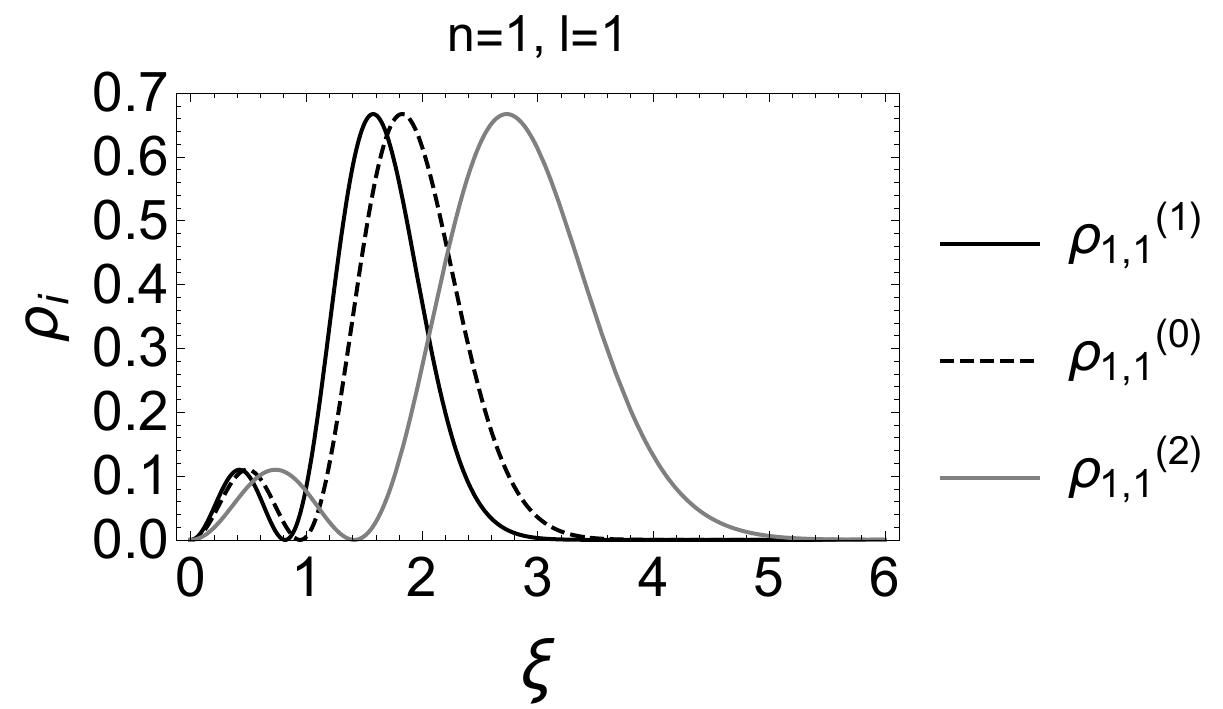}
\end{minipage}\\
\begin{minipage}[h]{0.8\linewidth}
\includegraphics[width=1\linewidth]{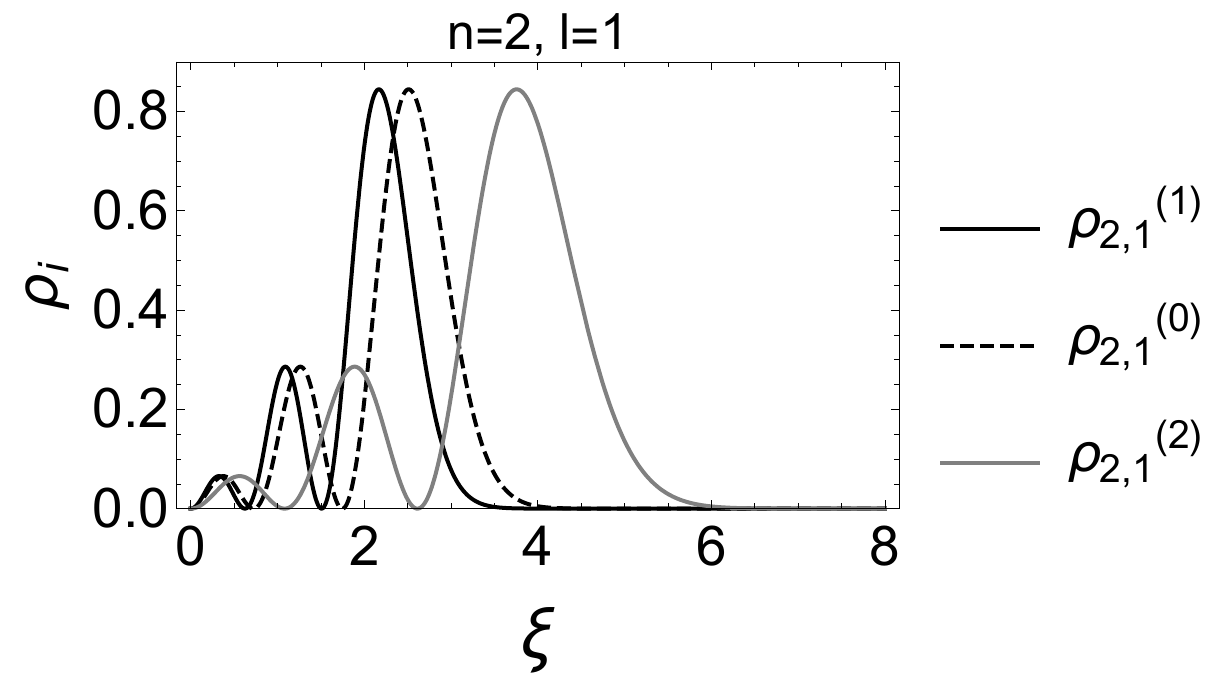}
\end{minipage}
\caption{\label{fig:eigenfunctions}
Probability distributions of the $l=1$ eigenstates (\ref{splitting-DKP-PDF}) for
(a) $n=0$ (top), (b) $n=1$ (center) and (c) $n=2$ (bottom)
with $\omega=2\widetilde{\omega}=\frac{qB}{2mc}>0$ and $\xi=r/a$.
The splitting implies the case $-\omega=2\widetilde{\omega}<0$ is obtained by exchanging $s_1$ and $s_2$
($\rho_1\leftrightarrow\rho_2$), which means
to reverse the direction of $\mathbf{B}$.}
\end{figure}

\section*{Analysis of the results}
Now we analyze the energies and the eigenstates probability distributions.
For fixing ideas we set
$2|\widetilde{\omega}|=\omega$
and $\frac{m\omega a^2}{\hbar}=1$ so we have
$\omega_1=\omega(1+\frac{1}{2}\textrm{sign}(\widetilde{\omega}))$,
$\omega_0=\frac{\sqrt{5}}{2}\omega$, and
$\omega_2=\omega(1-\frac{1}{2}\textrm{sign}(\widetilde{\omega}))$.
If we define $\alpha_i=\frac{m\omega_ia^2}{\hbar}$ it follows that
$\alpha_1=1+\frac{1}{2}\textrm{sign}(\widetilde{\omega})$,
$\alpha_0=\sqrt{5}/2$ and $\alpha_2=1-\frac{1}{2}\textrm{sign}(\widetilde{\omega})$,
where $\textrm{sign}(\gamma)$ denotes the sign function of $\gamma$ for all $\gamma\in\mathbf{R}$.
Their corresponding probability distributions are obtained from the squared modulus of
(\ref{scalar-DKPO-solution}) and (\ref{solution-v}), expressed in the compact form by
\bey
\rho_{n,l}^{(i)}(\xi)=\left|\alpha_i \xi^2\right|^{|l|}e^{-\alpha_i \xi^2} L^{2|l|}_n \left(\alpha_i \xi^2\right),
\label{splitting-DKP-PDF}
\eey
with $i=0,1,2$, $\xi=r/a$ a dimensionless variable and $a$ a characteristic length
(whose meaning will be clear later). The
dependence in $\theta$ has disappeared due to the spherical symmetry of the problem.
In order to study the interplay between the splitting and the energies, from (\ref{scalar-DKPO-spectrum})
and (\ref{spectrum-v}) we can recast the energies for the splitting as
\ben
&(\epsilon_{n,l}^{(i)})^2=\nonumber\\
&1-\frac{2\hbar(|l|\widetilde{\omega}+\omega)}{mc^2}\delta_{0i}
+\frac{\hbar \omega_i}{mc^2}\left(4n+2|l|\left(1-s_i\right)+2(s_i+\delta_{0i})\right), \label{splitting-spectrum}
\een
where $\epsilon_{n,l}^{i}$ are the energies adimensionalized by $mc^2$, $\delta_{0i}$ the Kronecker delta, $i=0,1,2$ and
$s_0=0, s_1=-s_2=1$.
\begin{figure}
\centering
\begin{minipage}[h]{0.8\linewidth}
\includegraphics[width=\linewidth]{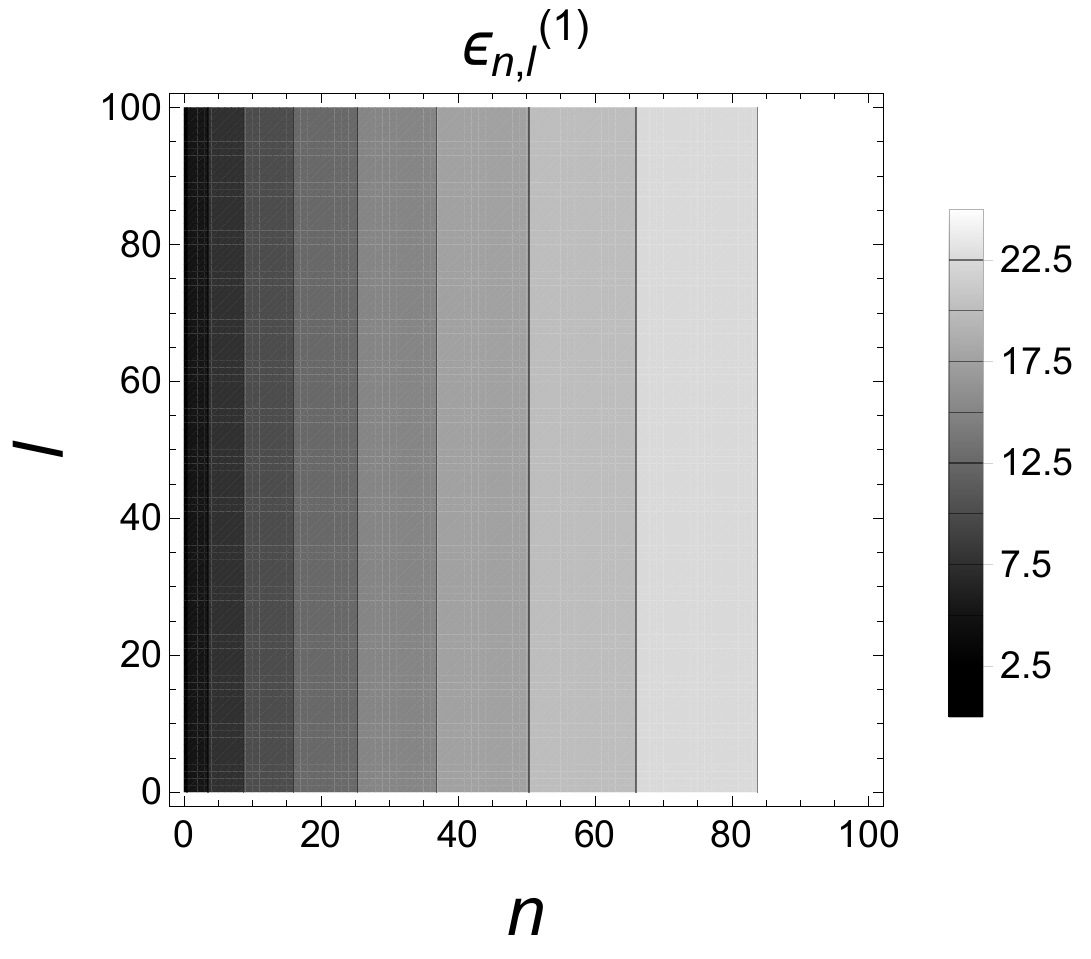}
\end{minipage}\\
\begin{minipage}[h]{0.8\linewidth}
\includegraphics[width=\linewidth]{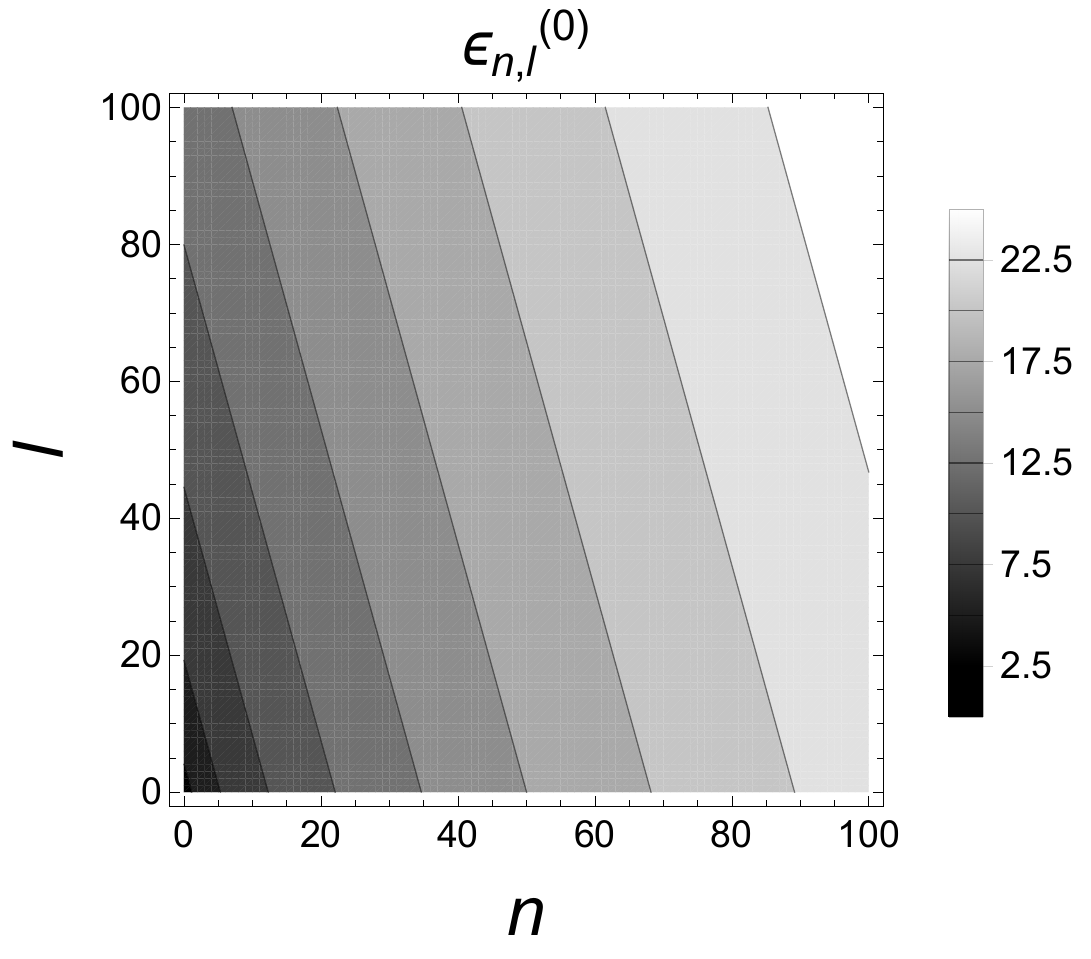}
\end{minipage}\\
\begin{minipage}[h]{0.8\linewidth}
\includegraphics[width=\linewidth]{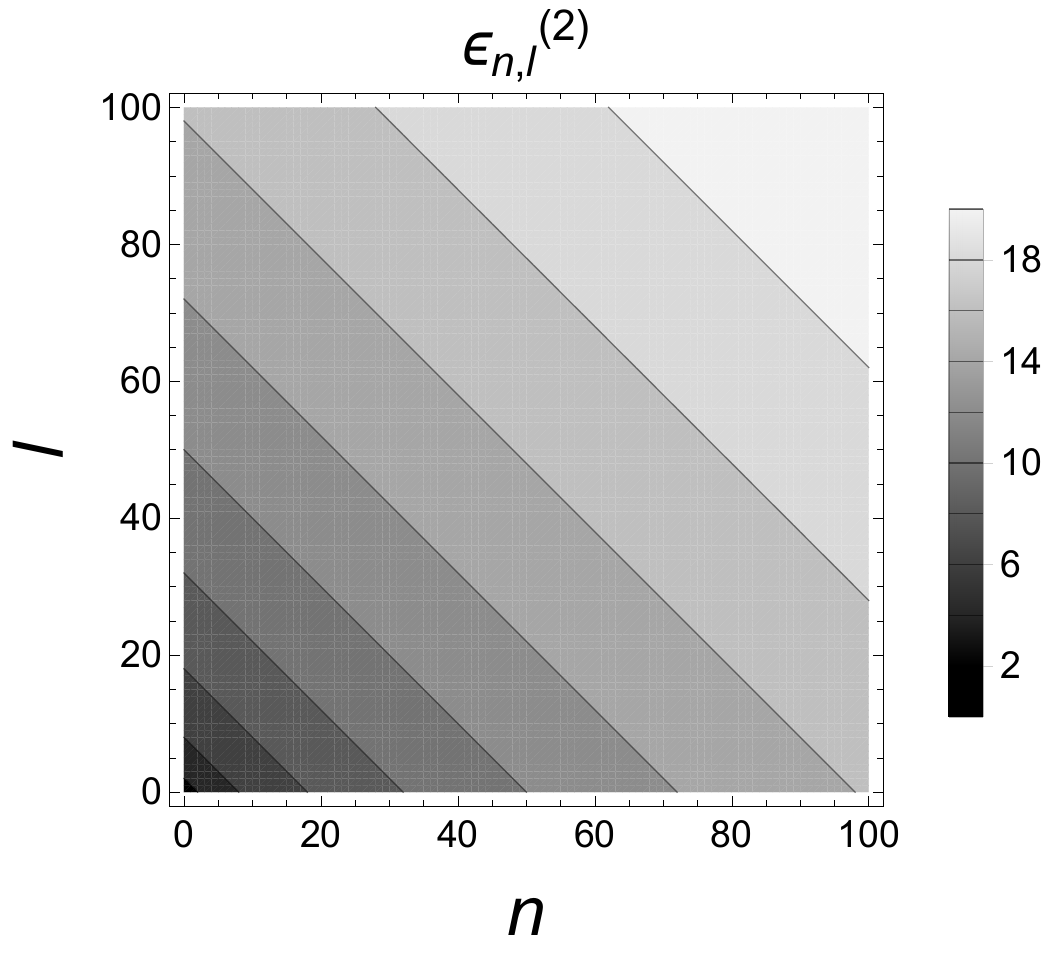}
\end{minipage}
\caption{\label{fig:eigenergies}
Contour plots of the splitting energy levels (\ref{splitting-spectrum-dimensionless}): (a) $s_1=1$ (top),
(b) $s_0=0$ (center) and (c) $s_2=-1$ (bottom) for
$\omega=2\widetilde{\omega}=\frac{qB}{2mc}>0$ within the range $0\leq n,l\leq100$, .
Due to the symmetry of (\ref{splitting-spectrum-dimensionless})
around $l=0$, i.e. $\epsilon_{n,l}^{(i)}=\epsilon_{n,-l}^{(i)}$, only the positive range of
$l$ is shown. The lines indicate the degeneracy
whose slopes are, from \eqref{slopes}, $\infty$, $-\frac{2}{1-\frac{1}{\sqrt{5}}}$ and
$-1$ for $s_1=1$, $s_0=0$, and $s_2=-1$.}
\end{figure}
For compatibilizing with $\frac{m\omega a^2}{\hbar}=1$ we set $\frac{\hbar\omega}{mc^2}=1$
so the characteristic length $a$ results to be the Compton wavelength $a=\frac{\hbar}{mc}$, which is a
natural representation for mass on the quantum scale.
The formula (\ref{splitting-spectrum}) turns out
\ben
&(\epsilon_{n,l}^{(i)})^2=1 \nonumber\\
&-(|l|\textrm{sign}(\widetilde{\omega})+2)\delta_{0i}
+\alpha_i\left(4n+2|l|\left(1-s_i\right)+2(s_i+\delta_{0i})\right). \label{splitting-spectrum-dimensionless}
\een
In Fig. \ref{fig:eigenfunctions} we show the probability distributions (\ref{splitting-DKP-PDF})
of the $l=1$ eigenstates $\varphi_i$ with $i=0,1,2$ for $n=0,1,2$ and $\widetilde{\omega}>0$. We can see that
the distributions of the spin projections $s_1=1$ and
$s_0=0$ are close while the corresponding to $s_2=-1$ exhibits a different behavior,
as a consequence of the splitting frequencies.
In Fig. \ref{fig:eigenergies} we show the energies (\ref{splitting-spectrum-dimensionless})
of the eigenstates $\varphi_i$ for $0\leq n,l\leq100$ and $\widetilde{\omega}>0$.
The straight lines of Fig. \ref{fig:eigenergies} correspond to the degenerated states
whose slopes
are only depending on the sign of $\widetilde{\omega}$, that is, they are in function of the
sign of $\mathbf{B}$. In fact, by letting $\varepsilon_{n,l}^{(i)}=k_i=\textrm{constant}$ in
\eqref{splitting-spectrum}
the curves $l=l(n)$ of degeneracy in the range $l\geq0$ have the slopes
\bey\label{slopes}
-\frac{2}{1-\frac{\widetilde{\omega}}{\sqrt{\omega^2+\widetilde{\omega}^2}}}
\quad, && \quad \textrm{for} \quad s_0=0  \nonumber    \\
-\frac{2}{1-s_i}  \quad, && \quad \textrm{for} \quad s_1=-s_2=1
\eey
where $\omega_{1,2}\neq0$.
From (\ref{splitting-spectrum}) some more features about the energies can be highlighted.
For the component
of projection $s_1=1$ the vibrational ($l=0$) and rotational ($l\neq0$) states
with quantum number $n$ have identical energy, so the magnetic field turns off the angular momentum
and spin effects.
It is also worth noting that in virtue of the splitting,
the case $\widetilde{\omega}<0$ has the same eigenfunctions and energies but with the
components flipped $\varphi_1\rightarrow\varphi_2$, while $\varphi_0$ remains invariant.

\section*{Thermodynamics of the vectorial sector}

\begin{figure}[bt]
\centering
\begin{minipage}[h]{1\linewidth}
\includegraphics[width=\linewidth]{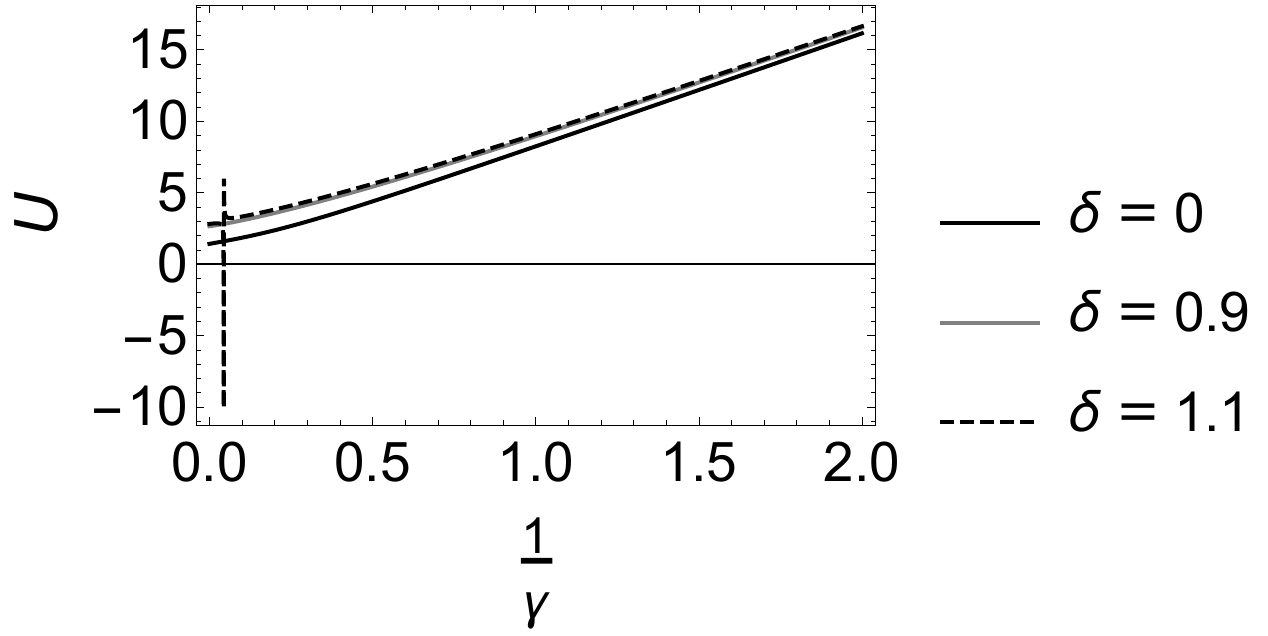}
\end{minipage}\\
\begin{minipage}[h]{1\linewidth}
\includegraphics[width=\linewidth]{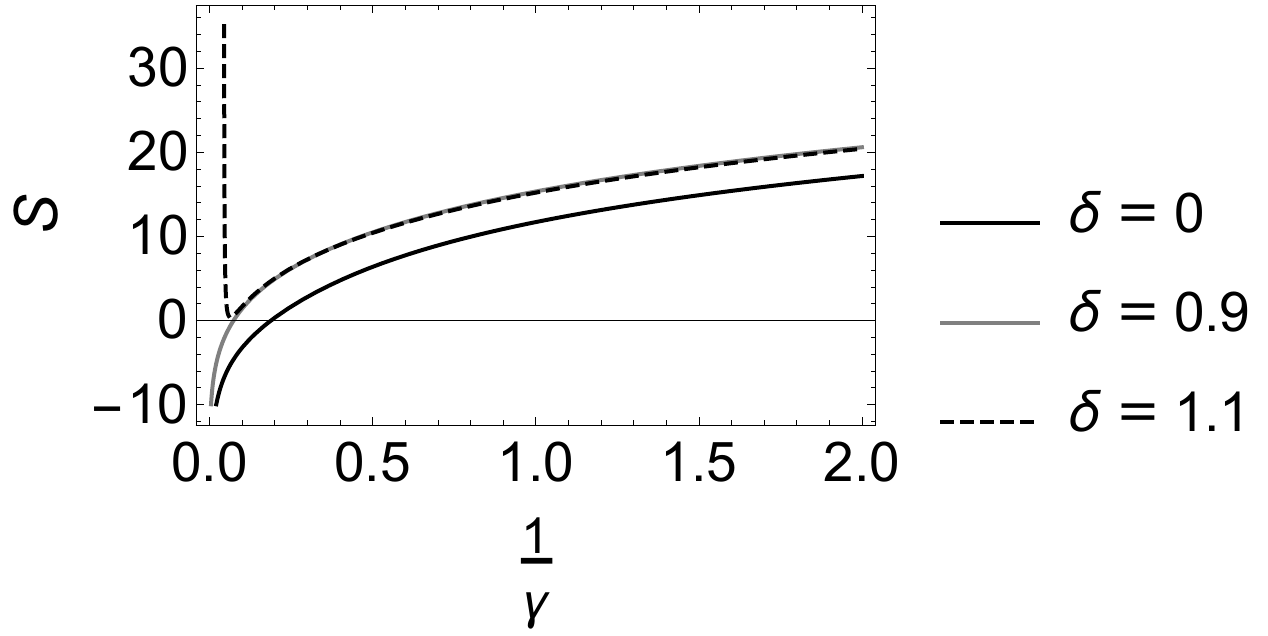}
\end{minipage}\\
\begin{minipage}[h]{0.9\linewidth}
\includegraphics[width=\linewidth]{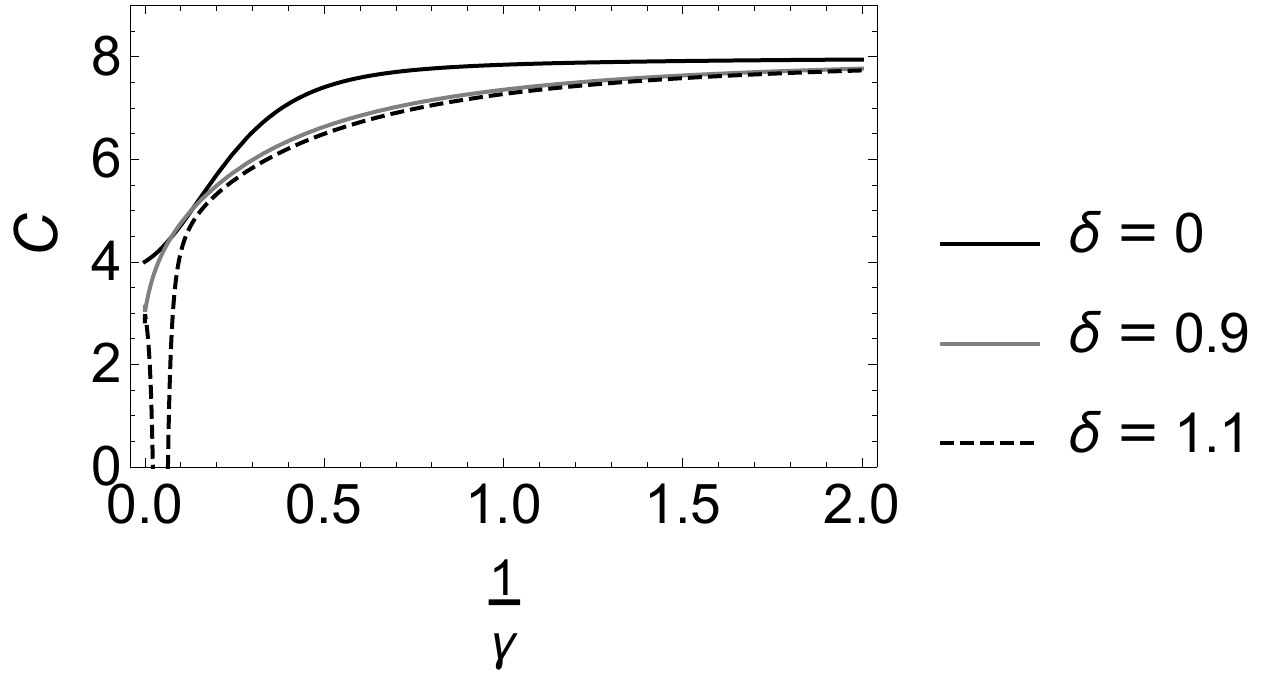}
\end{minipage}
\caption{\label{fig:thermodynamical-potentials}
(a) Internal energy (top), (b) entropy (center) and (d) specific heat (bottom)
of the vectorial sector of the two-dimensional DKP oscillator in an external magnetic field
$\mathbf{B}=B\hat{z}$ in function
of the real parameter $\delta\in[0,\infty)$, with $\delta=\frac{qB}{mc\omega}$
and $\omega=\frac{mc^2}{2\hbar}$ the oscillator frequency.
The discontinuity of the dashed curves indicate the phase transition that occurs
around $\delta=1$ due to the cancellation of the
oscillation of the 2-component.}
\end{figure}

We illustrate the effect of the splitting in the statistical properties of the vectorial sector
of the DKP oscillator since the scalar sector does not exhibit a
cancellation of the oscillation.
For accomplish this, we consider that the system is at
equilibrium with a thermal bath of finite temperature $T$ in order to obtain the partition
function of its canonical ensemble. Also,
we consider only the states with positive energy due to those with negative energy
are unlimited below, thus ensuring a stable ensemble \cite{Pacheco-EPL}.
Then, the partition function results
\bey\label{partition-function}
Z=\left(\sum_{n_1,l_1}\exp\left(-\gamma y_{n_1,l_1}\right)\right)
\times\left(\sum_{n_2,l_2}\exp\left(-\gamma y_{n_2,l_2}\right)\right)
\eey
with $\gamma=\frac{mc^2}{k_BT}$ and
\bey\label{partition-constants}
y_{n_i,l_i}^2=
1+\alpha_i\left(4n_i+2|l_i|\left(1-s_i\right)+2s_i)\right) \quad, \quad i=1,2 .
\eey
Here $k_B$ is the Boltzmann constant and
we have employed again
$\alpha_i=\frac{\hbar\omega_i}{mc^2}$ with $\omega_i=\omega+s_i\widetilde{\omega}$ for $i=1,2$.
For facilitating the calculations we make $\frac{\hbar\omega}{mc^2}=1/2$ and then $\alpha_i=
\frac{1}{2}(1+s_i\delta)$
with $\widetilde{\omega}=\delta\omega$.
We also consider $\delta\in[0,\infty)$ to avoid negative expressions
in the right hand of \eqref{partition-constants}
and thus to simplify the counting of degeneracies.
Since $y_{n_1,l_1}^2=2(1+\delta)n_1+2+\delta$
and $y_{n_2,l_2}^2=2(1-\delta)(n_2+|l_2|)+\delta$ we can calculate
the sums of \eqref{partition-function} separately by means of the general formula
\bey \label{general-formula}
\sum_{k}\Omega(E_k)\exp\left(-\beta E_k\right),
\eey
where $\Omega(E_k)$ indicates the degeneracy of the energy level $E_k$. For avoiding
the infinite degeneracy of $y_{n_1,l_1}$ in $l_1$ we assume that
$\Omega(E_{n_1,l_1})=\sum_{n_1=-l}^{n_1}=2n_1$, which physically means that only the
states with $|l_1|\leq n_1$ contribute significatively to $E_{n_1,l_1}$
and the rest of terms can be neglected.
Then, we have
\bey\label{sum-partition-function-1}
\sum_{n_1,l_1}\exp\left(-\gamma y_{n_1,l_1}\right)&=&
\sum_{n=0}^{\infty}\sum_{l=-n}^{n}
\exp\left(-\gamma\sqrt{2(1+\delta)n+2+\delta}\right)\nonumber\\
&=&2\sum_{n=0}^{\infty} n\exp\left(-\gamma\sqrt{2(1+\delta)n+2+\delta}\right).
\eey
For $y_{n_2,l_2}$ we have a double degeneracy given by all the pairs $(n_2,l_2)$ such
that $n_2+l_2=k\in\mathbf{N}_0$ with $n_2,l_2=0,\ldots,k$. This implies that
$\Omega(E_{n_2,l_2})=2\sum^{k+1}1=2(k+1)$ for $\beta E_{n_2,l_2}=k$. Then, we have
\bey\label{sum-partition-function-2}
&\sum_{n_2,l_2}\exp\left(-\gamma y_{n_2,l_2}\right)=\nonumber\\
&2\sum_{n=0}^{\infty} (n+1)\exp\left(-\gamma\sqrt{2(1-\delta)n+\delta}\right).
\eey
The sums \eqref{sum-partition-function-1} and \eqref{sum-partition-function-2}
can be calculated with the help of the Euler-Maclaurin's formula
(employed in relativistic
contexts for instance in \cite{Pacheco-EPL,Nouicer})
\bey\label{Euler-Maclaurin}
\sum_{n=0}^{\infty}f(n)=\frac{1}{2}f(0)+\int_{0}^{\infty}f(x)dx-\sum_{p=1}^{\infty}\frac{1}{(2p)!}B_{2p}f^{2p-1}(0),
\eey
with $B_{2p}$ the Bernoulli's numbers and $f^{2p-1}(0)$ the derivatives of odd order of $f(x)$ at $x=0$.
Now the crucial observation is that in the limit of high thermal excitations $\gamma=\frac{mc^2}{k_bT}\ll1$ since
the first and third terms of \eqref{Euler-Maclaurin} only have powers of $\gamma$ and the integral
has powers of $\gamma^{-1}$, then it is enough to consider only the integral , i.e.
\bey\label{partition-function-integral}
&Z\approx\left(2\int_{0}^{\infty}x\exp\left(-\gamma\sqrt{2(1+\delta)x+2+\delta}\right)dx\right)\times\nonumber\\
&\left(2\int_{0}^{\infty}(x+1)\exp\left(-\gamma\sqrt{2(1-\delta)x+\delta}\right)dx\right) \quad \textrm{for} \ \gamma\ll1.
\eey
Using in \eqref{partition-function-integral} that for $a,b,\gamma\geq0$
\bey\label{integral-formulas}
\int_{0}^{\infty}x\exp\left(-\gamma\sqrt{ax+b}\right)dx
=\frac{4 e^{-\gamma\sqrt{b}} \left(b \gamma ^2+3 \sqrt{b} \gamma +3\right)}{a^2 \gamma ^4}
\nonumber\\
\int_{0}^{\infty}(x+1)\exp\left(-\gamma\sqrt{ax+b}\right)dx=\nonumber\\
\frac{2 e^{-\gamma\sqrt{b}} \left(\sqrt{b} \gamma  \left(a \gamma ^2+6\right)+a \gamma ^2+2 b
\gamma ^2+6\right)}{a^2 \gamma ^4},
\eey
we obtain the partition function of the vectorial sector for $\gamma\ll1$
\bey\label{partition-function-vector-sector}
Z(\gamma,\delta)=\left(\frac{2e^{-\gamma\sqrt{2+\delta}} \left((2+\delta) \gamma ^2+3 \sqrt{2+\delta}
\gamma +3\right)}{(1+\delta)^2 \gamma ^4}\right)\times \nonumber\\
\left(\frac{e^{-\gamma\sqrt{\delta}} \left(\sqrt{\delta} \gamma  \left((1-\delta) \gamma ^2+6\right)+(1-\delta)
\gamma ^2+2 \delta
\gamma ^2+6\right)}{(1-\delta)^2 \gamma ^4}\right).
\eey
Thus, the partition function
\eqref{partition-function-vector-sector} results expressed in function of
$\delta$ and $\gamma=\frac{mc^2}{k_BT}$, with the latter measuring the ratio
between the rest mass and the
thermal energy.
It is worth noting that $Z(\gamma,\delta)\neq Z(\gamma,-\delta)$ due to the spectrum
of $\varphi_1$ and $\varphi_2$ is not flipped when $\delta\rightarrow-\delta$.
In the high temperature limit $T\rightarrow\infty$
we have $Z(\gamma,\delta)\approx \frac{6}{(1+\delta)^2\gamma^4}\frac{6}{(1-\delta)^2\gamma^4}$ and then
the partition function results symmetric in $\delta$.
On the other hand, the partition function \eqref{partition-function-vector-sector}
is real for $0\leq\delta \leq 1$ and when $\delta>1$ results complex,
presenting divergences in all the interval $(1,\infty)$.
In view of \eqref{integral-formulas} we can see that
the individual
partition functions of $\varphi_1$ and $\varphi_2$ have the same structure $\propto 1/((1+s_i\delta)^2\gamma^4)$, thus
expressing that
only the component $\varphi_2$ experiments a phase transition when $\delta\rightarrow1$.
This corresponds
to the particle free case $\widetilde{\omega}=\omega$ with $\omega_2\rightarrow0$, as pointed out previously.
From the thermodynamic relations
in function of the dimensionless variable $\gamma$ \cite{Pacheco-EPL}
\bey\label{thermodynamical-relations}
U &=& -mc^2\frac{\partial \ln Z}{\partial\gamma} \quad \textrm{(internal energy)} \nonumber\\
F &=& -mc^2\frac{1}{\gamma}\ln Z  \quad \textrm{(Helmholtz free energy)} \nonumber\\
S &=& k_B\gamma^2\frac{\partial F}{\partial \gamma}  \quad \textrm{(entropy)} \nonumber\\
C &=& -k_B\gamma^2\frac{\partial U}{\partial \gamma}  \quad \textrm{(specific heat)},
\eey
we derive the thermodynamics of the vectorial sector of the DKP oscillator. Substituting
\eqref{partition-function-vector-sector} in \eqref{thermodynamical-relations}
we illustrate in Fig. \ref{fig:thermodynamical-potentials}
the internal energy, the entropy and the specific heat.
We can see that the thermodynamic potentials rapidly converge to their asymptotical
behaviors when $1/\gamma\rightarrow\infty (T\rightarrow\infty)$,
in virtue of that
the thermal excitations
erase the particularities of the vectorial sector spectrum as soon as
the only relevant term in
the total partition function \eqref{partition-function-vector-sector} is proportional to $T^8$.
From \eqref{partition-function-vector-sector}
the asymptotical expressions of
\eqref{thermodynamical-relations}, that also
show the additivity property in the vectorial sector, are given by
\bey\label{statistical-functions}
U \approx 8k_BT = 4k_BT + 4k_BT = U_1+U_2  \nonumber\\
F \approx -8k_BT\ln\left(\frac{k_BT}{mc^2(1-\delta^2)^{1/4}}\right)= \nonumber\\
-4k_BT\ln\left(\frac{k_BT}{mc^2(1+\delta)^{1/4}}\right)
- 4k_BT\ln\left(\frac{k_BT}{mc^2(1-\delta)^{1/4}}\right) \nonumber\\
= F_1+F_2 \nonumber\\
S \approx 8k_B\left(\ln\left(\frac{k_BT}{mc^2(1-\delta^2)^{1/4}}\right)+1\right)= \nonumber\\
4k_B\left(\ln\left(\frac{k_BT}{mc^2(1+\delta)^{1/4}}\right)+1\right)+
4k_B\left(\ln\left(\frac{k_BT}{mc^2(1-\delta)^{1/4}}\right)+1\right)\nonumber\\
C \approx 8k_B = 4k_B + 4k_B = C_1+C_2.
\eey

\section*{Conclusions}

We have revisited the $(2+1)$-dimensional DKP oscillator in an external magnetic field from
scalar $4\times 4$ and vectorial
$6\times 6$ representations, which allow to study several cases
of the literature as well as calculating their energies and eigenfunctions, in a unified way.
The energies and eigenfunctions
for the scalar DKPO in a uniform magnetic field
are
shown in the equations (\ref{scalar-DKPO-spectrum})-(\ref{scalar-DKPO-solution}), respectively, where
the angular frequency $\omega$ is
replaced by $\omega_0=\sqrt{\omega^2+\widetilde{\omega}^2}$ being $\widetilde{\omega}=\frac{qB}{mc}$ the field angular frequency.
The vector DKPO interacting with a magnetic field presents a splitting $\omega\rightarrow(\omega_1,\omega_0,\omega_2)$
in the frequency of the oscillation (Fig. \ref{fig:diagram}),
that corresponds to the spin projections $s_i=+1,0,-1$ of the vector DKP field.
The energies and the eigenfunctions of this oscillator are presented in
(\ref{scalar-DKPO-spectrum}), (\ref{scalar-DKPO-solution}), (\ref{spectrum-v}),
(\ref{solution-v}) and some of them illustrated in
Figs. \ref{fig:eigenfunctions}, \ref{fig:eigenergies}, where the slopes of the degeneracies \eqref{slopes} are
in function of $\omega$, $\widetilde{\omega}$ and $s_i$.
Some special cases have been studied with two
critical ones when $\widetilde{\omega}=\pm\omega\rightarrow \mathbf{B}=\pm\frac{2mc\omega}{q}\hat{z}$.
In these cases the component $\varphi_1$ or $\varphi_2$ stops oscillating, where
the phase transition $\widetilde{\omega}\rightarrow\omega$ has been characterized by means of the canonical ensemble
of the vectorial sector. The thermodynamic potentials exhibit a rapid convergence to their
asymptotical expressions due to the thermal excitations (Fig. \ref{fig:thermodynamical-potentials}), thus
resulting the partition function \eqref{partition-function-vector-sector} $\propto T^8$ and the individual ones $\propto T^4$, with
the divergences
associated to the critical values
of the magnetic field.
\section*{Acknowledgments}
	The authors acknowledge support received from the National Institute of Science
	and Technology for Complex Systems (INCT-SC),
	and from the CNPq and the CAPES (Brazilian agencies) at Universidade Federal da Bahia, Brazil.


\end{document}